\documentclass[journal = jpclcd, manuscript = article, layout = twocolumn]{achemso}

\usepackage{newtxtext}
\usepackage{newtxmath}
\usepackage[scaled=0.9]{helvet}
\usepackage[fontsize=9pt]{fontsize}
\usepackage[version = 3]{mhchem}
\usepackage{graphicx}
\usepackage{amsmath}
\usepackage{amsfonts}
\usepackage{xcolor}

\newcommand{\tb}[1]{\textbf{#1}}

\renewcommand{\=}{\mbox{\,=\,}}
\newcommand{\e}{\mathrm{e}}
\renewcommand{\i}{\mathrm{i}}
\renewcommand{\d}{\mathrm{d}}

\title{Manipulating Two-Photon Absorption of Molecules through Efficient Optimization of Entangled Light}
\author{Sajal Kumar Giri}
\affiliation{Department of Chemistry, Northwestern University, 2145 Sheridan Road, Evanston, Illinois 60208, United States}
\author{George C. Schatz}
\email{g-schatz@northwestern.edu}
\affiliation{Department of Chemistry, Northwestern University, 2145 Sheridan Road, Evanston, Illinois 60208, United States}

\begin{document}
\begin{abstract} 
%Entangled light has generated a lot of interest lately as sources of two entangled photons are ideal in many respects for applications in sensing, low-intensity imaging of biomolecules, non-linear high-resolution spectroscopy, etc.
We report how the unique temporal and spectral features of pulsed entangled photons from a parametric downconversion source can be utilized for manipulating electronic excitations through the optimization of their spectral phase.
%We develop a new comprehensive optimization protocol based on Bayesian optimization for the two-photon state to selectively excite electronic states.
A new comprehensive optimization protocol based on Bayesian optimization has been developed in this work to selectively excite electronic states accessible by two-photon absorption.
Using our optimization method, the entangled two-photon absorption probability for a thiophene dendrimer can be enhanced by up to a factor of 20 while classical light turns out to be nonoptimizable.
Moreover, the optimization involving photon entanglement enables selective excitation that would not be possible otherwise.   
In addition to optimization, we have explored entangled two-photon absorption in the small entanglement time limit
showing that entangled light can excite molecular electronic states that are vanishingly small for classical light.
We demonstrate these opportunities with an application to a thiophene dendrimer.
\begin{tocentry}
    \centering
    \includegraphics[width = 1.0 \textwidth]{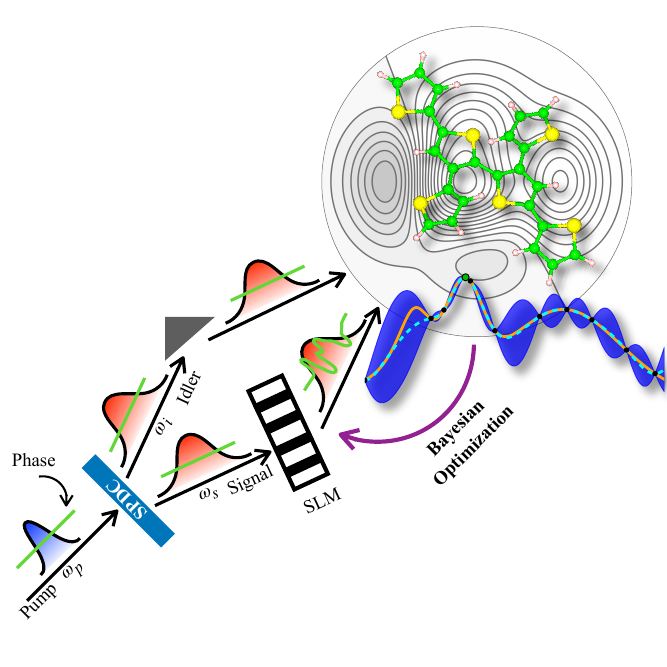}
\end{tocentry}
\end{abstract}

\maketitle

Two-photon absorption (TPA) has come a long way since 1931 when it was first predicted theoretically by G{\" o}ppert-Mayer et\,al.\cite{Mayer1931Uber} and verified experimentally by Kaiser et\,al.\cite{Kaiser1961Two} later in 1961 after the invention of optical lasers. 
It is a widely studied nonlinear process with potential applications where a molecule is excited by the absorption of two photons populating an electronic state with an energy equal to the sum of photon energies\cite{Guang1995Two,Albota1998Design,Belfield2000Multiphoton,Terenziani2008Enhanced,Dudek2020Two}.  
Here the photons arrive at random and the absorption rate is proportional to the instantaneous photon flux, $R_{r}\=\delta_{r}\Phi^{2}$, where $\delta_{r}$ is the classical TPA cross-section and $\Phi$ is the photon flux. 
As compared to one-photon absorption (OPA), TPA provides deeper penetration in the scattering media with increased focus and less bleaching\cite{Denk1990Two,Drobizhev2009Absolute,Drobizhev2011Two}. 
Despite these advantages, it has been observed that the TPA rate is extremely small for organic molecules. TPA may be enhanced in two ways,  either by considerably increasing the photon flux with CW excitation or by utilizing a broadband pulse.
Both of these have drawbacks: For the former situation, the intense light may considerably damage the sample while generating signals\cite{Casacio2021Nonclassical} and for the latter situation, the improvement in TPA rate is restricted by the spectral overlap between the light field and two-photon absorption.  
In the past, significant efforts have been put into the design of organic materials to obtain a large TPA cross-section\cite{Albota1998Design,Pawlicki2009Two} with minimal damage. \\

The spontaneous parametric down-conversion (SPDC) process\cite{Rubin1994Theory} generates photon pairs (historically known as signal and idler) typically in the optical regime with high temporal and spectral correlations which can induce coherent absorption of two photons when targeted to a photo-absorbing media. This is called entangled two-photon absorption (ETPA).
The photon correlations are characterized by an entanglement time $T_{e}$ which is the maximum time delay between the two entangled photons to be detected, i.e., after detecting one photon within this timescale the second photon must be detected. 
Each photon pair can initiate the two-photon process and the absorption rate increases linearly with the photon flux, $R_{e}\=\sigma_{e}\Phi$ where $\sigma_{e}$ is the ETPA cross-section. 
This linear dependence was first demonstrated experimentally with cesium (Cs) atoms\cite{Georgiades1995Nonclassical} and later with organic chromophores\cite{Lee2006Entangled}.
Due to the linear dependence, ETPA shows detectable excitation even at extremely low intensities where TPA signals are very weak with a quadratic dependence on $\Phi$.     
The high degree of temporal and spectral correlations of nonclassical light makes them unique from conventional light sources, especially for the study of two-photon absorption in molecules\cite{Lee2006Entangled,Juan2017Entangled,Burdick2018Predicting,Kang2020Efficient,Tabakaev2021Energy,Eshun2022Entangled,Samuel2022Experimental}. 
ETPA has several advantages over TPA including two-photon-induced transparency\cite{Fei1992Entanglement},  high-quality imaging of organic materials with low photon flux\cite{Varnavski2020Two,Varnavski2022Quantum}, enhanced fluorescence\cite{Upton2013Optically}, resonance energy transfer\cite{Avanaki2019Entangled}, and many more.   \\

The control of photoinduced processes using shaped classical light pulses is a well-studied area of research\cite{Meshulach1998Coherent,Rabitz2000Whither} with a large number of applications including chemical reactions\cite{Assion1998Control}, vibrational excitation in molecules\cite{Beumee1992Robust}, etc.  
Pulse shaping for two-photon and multi-photon absorption has already involved major theoretical and experimental advances\cite{Dudovich2001Transform,Silva2009Control,Fisher2018Pulse,Lahiri2021Controlling}  
where typically strong laser pulses control the phase of induced transitions leading to constructive or destructive interference between them.  
In 2004, Dayan et\,al.\cite{Dayan2004Two} experimentally demonstrated pulse shaping methods with broadband entangled light to control the two-photon absorption in rubidium (Rb) atom. 
It has been experimentally shown that the entangled two-photon state can be shaped by manipulating the spectral phase similar to a coherent ultrashort pulse\cite{Peer2005Temporal}.  
Typically a spatial light modulator (SLM) is used in the experiment in optimizing the spectral amplitude and phase of the ultrashort pulses\cite{Meshulach1998Coherent,Rabitz2000Whither,Dudovich2001Transform}.   
A recent study by Burdick et al.\cite{Burdick2021Enhancing} reported that the entanglement time and area can be modulated using a spectral filter where the large entanglement time limit has been used (without phase optimization) to maximize the ETPA cross-section. \\

In this letter, we investigate state-specific two-photon absorption of molecules, leveraging the unique properties of the photon pairs generated through the degenerate (equal photon frequencies) type-II SPDC process. 
Particularly we are interested in controlling the electronic excitation of molecules by optimizing the spectral phase of the pulsed entangled light and we will use this approach to explore the variation of ETPA enhancement with different intermediate and two-photon excited states.
As part of this, we develop a spectral phase optimization method for entangled light that selectively
manipulates two-photon excitations enabling both enhancement and selectivity in ETPA. 
In another part of this letter that is not related to phase optimization, we study the small entanglement time limit (the opposite limit compared to the study of Burdick et al.\cite{Burdick2021Enhancing}) where the ETPA cross-section no longer depends on the energies of intermediate states, and states that are normally dark in TPA are able to gain intensity in ETPA without optimization. \\
%We provide results in an application to the molecule T6, a thiophene dendrimer that has been the subject of unoptimized EPTA studies, and where theory and experiment are in good agreement. \\

We assume that the molecule is initially in its ground state $|g\rangle$ and it has multiple excited states $|e\rangle$ and $|f\rangle$ accessible through one- and two-photon transitions respectively. 
The two-photon transition amplitude to a final excited state $|f\rangle$ via intermediate states $|e\rangle$ is given by\cite{Tannoudji2011Advances} 
\begin{align}
    \label{tamp_eq}
    T_{f}^{ab}&\=\int_{-\infty}^{\infty}\d t_{2}\int_{-\infty}^{t_{2}}\d t_{1}M_{c}^{ab}(t_{1},\,t_{2})E_{c}(t_{1},\,t_{2}),  
\end{align}
with
\begin{align}
    \label{mc_eq}
    M_{c}^{ab}(t_{1},\,t_{2})&\=\sum_{e}D_{e}^{ab}\e^{\i(\varepsilon_{f}-\varepsilon_{e}-\i\gamma_{e}/2)t_{2}}\e^{\i(\varepsilon_{e}-\varepsilon_{g}+\i\gamma_{e}/2)t_{1}}, \\
    \label{ec_eq}
    E_{c}(t_{1},\,t_{2})&\=\langle 0|\hat{E}_{1}^{\dagger}(t_{1})\hat{E}_{2}^{\dagger}(t_{2})|\Psi\rangle \nonumber \\
    &\hspace{0.4cm}+\langle 0|\hat{E}_{2}^{\dagger}(t_{1})\hat{E}_{1}^{\dagger}(t_{2})|\Psi\rangle,
\end{align}
where $M_{c}^{ab}(t_{1},\,t_{2})$ is the dipole and $E_{c}(t_{1},\,t_{2})$ is the field correlation function.
Throughout we have used atomic units unless specified explicitly. 
At $t_{1}$ a transition occurs from $|g\rangle$ to $|e\rangle$ and at $t_{2}$ the $|\e\rangle$ state populations are transferred to $|f\rangle$.  
$D_{e}^{ab}$ are the two-photon transition matrix elements with two dipole-moment operators $\hat{\mu}_{a\in(x,\,y,\,z)}$ and $\hat{\mu}_{b\in(x,\,y,\,z)}$, $D_{e}^{ab}\=\langle f|\hat{\mu}_{a}|e\rangle\langle e|\hat{\mu}_{b}|g\rangle$.
The energies of $|g\rangle$, $|e\rangle$, and $|f\rangle$ states are denoted as $\varepsilon_{g}$, $\varepsilon_{e}$, and $\varepsilon_{f}$ respectively.  
A phenomenological line width $\gamma_{e}$ is introduced in Eq.\ref{mc_eq} for the intermediate state energy to account for dissipation that may emerge from various effects such as spontaneous emission, solvent effects, etc. This is important when the transition timescale is comparable to or smaller than the decay timescale.
We use two different $\gamma_{e}$ values, $\gamma_{e}\=0$ and $\gamma_{e}\=0.5$ eV, to overcome the difficulty of knowing the actual damping factor in the experiment. \\

Electric field operators in the field correlation function Eq.\ref{ec_eq} are given by
\begin{align}
    \label{field_eq}
    \hat{E}_{n}^{\dagger}(t)\=\sqrt{\frac{\hbar}{4\pi \epsilon_{0}cA}}\int_{0}^{\infty}\d\omega_{n}\sqrt{\omega_{n}}\hat{a}_{n}(\omega_{n})\e^{-\i\omega_{n}t}\e^{\i\phi(\omega_{n})},
\end{align}
where $a(\omega)$ ($a^{\dagger}(\omega)$) is the Bosonic annihilation (creation) operator of the photonic mode with a frequency $\omega$, satisfying the commutation relation $[\hat{a}_{1}(\omega_{1}),\,\hat{a}_{2}^{\dagger}(\omega_{2})]\=\delta(\omega_{1}-\omega_{2})\delta_{12}$.
The spectral phase is denoted by $\phi(\omega)$; through this function we optimize the two-photon state.
$A$ is the quantization area (effective area of the field interacting with the molecule), $c$ is the speed of light, and $\epsilon_{0}$ is the vacuum permittivity.  
Here we consider the wavelength much larger than the dimension of the target sample and drop the spatial dependence of light. \\

Using a narrow band approximation of the generated photons ($\sigma_{p}<<\bar{\omega}_{1,\,2}$) we calculate the transition amplitude as 
\begin{align}
    \label{tamp_qf_ana_eq}
    T_{f}^{ab}&\,\propto\,\sqrt{\frac{T_{p}\bar{\omega}_{1}\bar{\omega}_{2}}{T_{e}}}\e^{-\frac{T_{p}^{2}}{8\ln 2}(\varepsilon_{f}-\varepsilon_{g}-\omega_{p})^{2}} \nonumber \\
    & \hspace{0.6cm} \sum_{e}D_{e}^{ab}\Bigg{[}\frac{1-\e^{-\gamma_{e}T_{e}/2}\e^{-\i\Delta_{e}^{(1)}T_{e}}}{\Delta_{e}^{(1)}-\i\gamma_{e}/2} \nonumber \\
    & \hspace{2cm} +\frac{1-\e^{-\gamma_{e}T_{e}/2}\e^{-\i\Delta_{e}^{(2)}T_{e}}}{\Delta_{e}^{(2)}-\i\gamma_{e}/2}\Bigg{]}, 
\end{align}
with detuning of the intermediate states $\Delta_{e}^{(1,\,2)}\=\varepsilon_{e}-\varepsilon_{g}-\bar{\omega}_{1,\,2}$.
In obtaining the above expression for the transition amplitude we have used $\phi(\omega)=0$ but we also consider arbitrary forms of $\phi(\omega)$ during its optimization.    
The pump spectral width $\sigma_{p}$ is connected with the pulse duration $T_{p}$ as $\sigma_{p}=4\ln2/T_{p}$.
We refer to the Supporting Information (SI) for the details of the derivation. 
The transition amplitude for classical light analogous to Eq.\ref{tamp_qf_ana_eq} is obtained as
\begin{align}
    \label{tamp_cf_ana_eq}
    \widetilde{T}_{f}^{ab}&\,\propto\,\sqrt{\bar{\omega}_{1}\bar{\omega}_{2}}\e^{-\frac{T_{p}^{2}}{16\ln 2}(\varepsilon_{f}-\varepsilon_{g}-\omega_{p})^{2}} \nonumber \\
    &\hspace{0.6cm}\sum_{e}D_{e}^{ab}\Bigg{[}\frac{1}{\Delta_{e}^{(1)}-\i\gamma_{e}/2}+\frac{1}{\Delta_{e}^{(2)}-\i\gamma_{e}/2}\Bigg{]}.
\end{align} \\

At room temperature molecules can rotate freely with respect to the laser polarization and therefore we perform rotational averaging of molecules in calculating the target state population\cite{Monson1970Polarization} 
\begin{align}
    \label{popf_eq}
    P_{f}\=\frac{1}{30}\sum_{a,\,b}\Big{[}FT_{f}^{ab}\bar{T}_{f}^{ab}+GT_{f}^{ab}\bar{T}_{f}^{ba}+ HT_{f}^{aa}\bar{T}_{f}^{bb}\Big{]},
\end{align}
with F$\=$G$\=$H$\=$2 for the parallel polarization and F$\=H\=$-1, G$\=$4 for the perpendicular polarization. 
Here $\bar{T}$ stands for the complex conjugate of $T$.
We have employed perpendicular (parallel) polarization for quantum (classical) light.
The (entangled) two-photon absorption spectrum is obtained from $P_{f}$, $S_{\rm (E)TPA}\=\sum_{f}P_{f}$. 
Note that we are interested in calculating transition probability, rather than a rate, and thus do not consider issues about state densities that were considered in the previous work by Kang et\,al\cite{Kang2020Efficient}.   \\

The main difference between quantum and classical amplitudes lies in the presence of a factor $\eta\=1-\e^{-\gamma_{e}T_{e}/2}\e^{-\i\Delta_{e}^{(1,\,2)}T_{e}}$ in the former one which modifies the transition amplitude to a specific intermediate state with a suitable choice of $T_{e}$. 
For a large entanglement time, $T_{e}\rightarrow\infty$, this factor is 1 and becomes identical to the classical result.  \\

A smaller entanglement time must be realized to observe the quantum effects in the transitions.
In order to do so we take the limit $T_{e}\rightarrow 0$ where the phase factor $\eta$ approximately becomes proportional to $1-\i T_{e}[\Delta_{e}-\i\gamma_{e}/2]$ and Eq.\ref{tamp_qf_ana_eq} reads 
\begin{align}
    \label{tamp_qf_ana_te0_eq}
    T_{f}^{ab}\,\propto\,\sqrt{T_{p}T_{e}\bar{\omega}_{1}\bar{\omega}_{2}}\e^{-\frac{T_{p}^{2}}{8\ln 2}(\varepsilon_{f}-\varepsilon_{g}-\omega_{p})^{2}}\sum_{e}D_{e}^{ab}.
\end{align}
Now interestingly in the above equation, we do not have any $\Delta_{e}$ terms and the transition becomes independent of the intermediate state energy $\varepsilon_{e}$. 
Note that if we have a complete set of intermediate states, the sum in Eq.\ref{tamp_qf_ana_te0_eq} reduces to the quadrupole matrix element $\langle f|\hat{\mu}_{a}\hat{\mu}_{b}|g\rangle$, which provides further simplification in the limit of short entanglement times.
This is significant, as it provides a mechanism for highly nonresonant intermediate states to contribute to the ETPA spectrum. 
We discuss this later in detail, as it leads to final states that do not contribute to TPA, or even to ETPA for larger $T_{e}$.  \\

We have performed electronic structure calculations with NWChem for the T6 thiophene dendrimer molecule to determine excited states and transition moments that will be used to illustrate how the theory works. 
As part of this we optimized the molecular structure using the B3LYP exchange correlation and def2-TZVP basis function.
Ten excited states have been considered to calculate energies and the transition dipole matrix using a recently developed method called second linear response time-dependent density functional theory (SLR-TDDFT)\cite{Mosquera2016Sequential,Mosquera2017Exciton,Mosquera2021Second}.
This method was developed to extend transitions involving the ground state under standard linear response TDDFT to transitions between the excited states. 
It has been used to successfully predict the TPA and ETPA cross-section of thiophene dendrimers\cite{Kang2020Efficient}. \\

\begin{figure}
    \centering
    \includegraphics[width = 0.35 \textwidth]{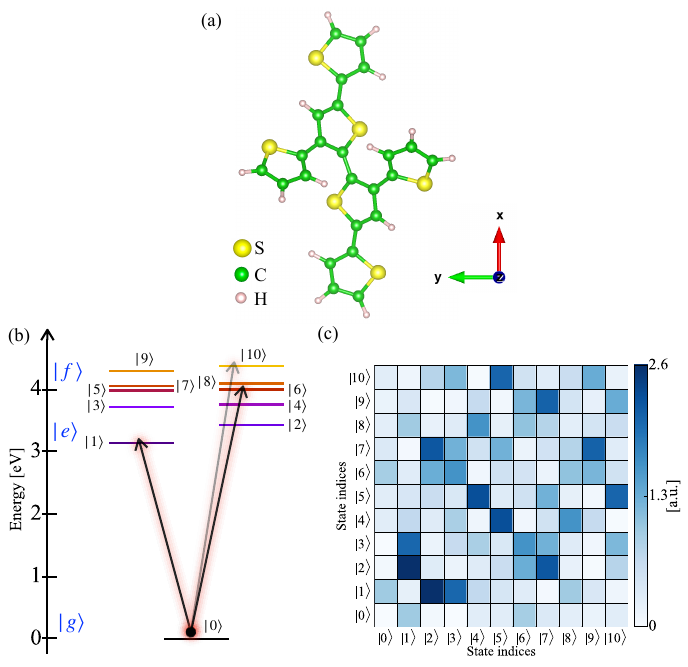}
    \caption{(a) Optimized geometry, (b) energy levels, and (c) transition dipole matrix of 6-polythiophene dendrimer molecule (T6).
    In panel (b), efficient one-photon transition paths are shown by black arrows with transition dipoles proportional to the line brightness.}  
   \label{sys_fig}
\end{figure}

\begin{figure}[h]
    \centering
    \includegraphics[width = 0.4 \textwidth]{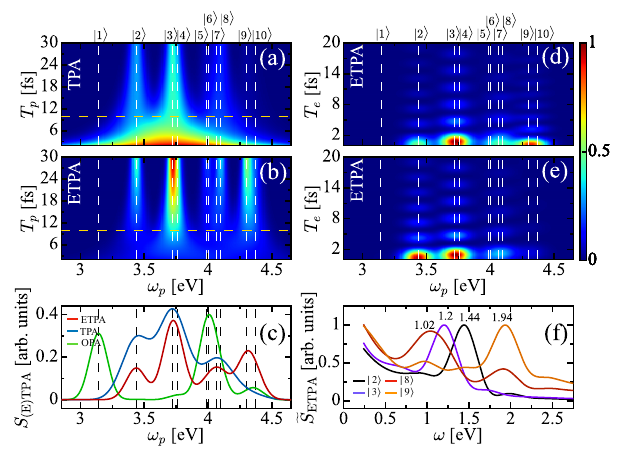}
    \caption{(a) Two-photon absorption (TPA) and (b) entangled two-photon absorption (ETPA) spectra as a function of $\omega_{p}$ and $T_{p}$. 
    Here, $T_{e}=1$ fs for the ETPA spectra.  
    (c) Spectra for $T_{p}=10$ fs ($\sigma_{p}=0.18$ eV) highlighted by dashed horizontal lines in (a), and (b). 
    The green line represents one-photon absorption (OPA).
    ETPA spectra for (d) all intermediate states and (e) intermediate states below 3.8 eV (where $|e\rangle$ state ranges in $|1\rangle$-$|4\rangle$) as a function of $\omega_{p}$ and $T_{e}$. 
    (f) Fourier transforms of $S_{\rm ETPA}$ for $|2\rangle$, $|3\rangle$, $|8\rangle$, and $|9\rangle$ excitation channels (from black to orange lines) where $\omega_{p}$ resonances with the $|f\rangle$ state energies.
    The dashed vertical lines show the excited state energies.
    }
   \label{t6_spec_wptpte_fig}
\end{figure}

In this work, we test our methods using the T6 thiophene dendrimer as a target molecule, but due to its generic nature, the general conclusions are also applicable to other organic chromophores.
The one- and two-photon absorption cross-sections of this molecule have been reported previously in experimental\cite{Harpham2009Thiophene,Badaeva2010Excited} and theoretical\cite{Badaeva2010Excited,Kang2020Efficient} works.
The DFT optimized geometry of the molecule is shown in Fig.\ref{sys_fig}(a), details of the T6 molecule and calculation of transition dipole moments are included in the SI. 
The excited states are indexed from $|1\rangle$ to $|10\rangle$ falling in the energy range of [3.14, 4.37] eV with the ground state energy $\varepsilon_{g}\=0$ as illustrated in Fig.\ref{sys_fig}(b).
Fig.\ref{sys_fig}(c) displays transition dipole matrix elements for these 10 excited states.
The ground state absorption spectrum of the optimized geometry is shown in Fig.\ref{t6_spec_wptpte_fig}(c) using a green solid line.
The calculated spectrum shows two major peaks at 3.14 eV ($\mu_{0,\,1}$ dipole matrix element) and 4 eV ($\mu_{0,\,6}$ dipole matrix element) which matches well with the experimentally measured peaks at 3.2 eV and 4.2 eV, respectively\cite{Harpham2009Thiophene}.  
There is also a small peak that appears at 4.37 eV corresponding to the state $|10\rangle$.
As a result, states $|1\rangle$, $|6\rangle$, and $|10\rangle$ are bright and the rest are dark for OPA. \\

We start our discussion with the study of ETPA in the small $T_{e}$ limit where we highlight important features that appear in ETPA compared to the corresponding TPA.  
To identify unique features of ETPA in the small $T_{e}$ limit, we have plotted TPA and ETPA spectra as a function of $\omega_{p}$ and $\sigma_{p}$ in Fig.\ref{t6_spec_wptpte_fig}(a) and Fig.\ref{t6_spec_wptpte_fig}(b) respectively. 
For ETPA spectra we used $T_{e}\=1$ fs.  
At sufficiently small $\sigma_{p}$, $\sigma_{p}\leq0.18$ eV ($T_{p}\geq10$ fs), three excitation channels are identifiable in TPA spectra and they appear at $3.14$ eV, $3.5$ eV, and $4.14$ eV.
The  $|f\rangle$ states involved are $|2\rangle$, $|3\rangle$, and $|8\rangle$ respectively, see the vertical lines and peak positions in Fig.\ref{t6_spec_wptpte_fig}(a).
Most interestingly, one additional peak at 4.3 eV appears in ETPA which is not seen in TPA. 
A closer look in Fig.\ref{t6_spec_wptpte_fig}(b) reveals that this excitation involves excitation of state $|9\rangle$.    
Note that, although OPA in Fig.\ref{t6_spec_wptpte_fig}(c) shows a little shoulder near the state $|10\rangle$, the state $|9\rangle$ is dark (though these states are very close in energy) whereas ETPA shows the additional peak.  
In Fig.\ref{t6_spec_wptpte_fig}(d) we see that all of the ETPA excitation channels show a decaying oscillation as a function of $T_{e}$ and the addition peak (state $|9\rangle$) is prominent when $T_{e}\rightarrow 0$. 
This extra peak which appears only in ETPA at the small $T_{e}$ limit is discussed in detail below. 
We have found that this observation is robust against the finite decay ($\gamma_{e}=0.5$ eV) of the intermediate states but its absolute amplitude goes down compared to that without including for decay. 
Therefore we conclude that the damping of the intermediate states does not have much influence in resolving the different excitation channels except for the absolute amplitude of the transitions. \\

To gain insight into this observation it is important to explore the transition paths (i.e., identifying all effective intermediate states) responsible for each excitation channel.
In OPA (Fig.\ref{t6_spec_wptpte_fig}(c)) we already have seen that there are three bright states $|1\rangle$, $|6\rangle$, and $|10\rangle$ available where the last peak is comparatively weak.
Therefore only these three states serve as intermediate states with the possible paths $|0\rangle\rightarrow|1\rangle\rightarrow|f\rangle$, $|0\rangle\rightarrow|6\rangle\rightarrow|f\rangle$, and $|0\rangle\rightarrow|10\rangle\rightarrow|f\rangle$, as shown by the black arrows in Fig.\ref{sys_fig}(c).
Before going further, note that we have $\Delta_{e}^{(1)}\=\Delta_{e}^{(2)}\=\Delta_{e}$ (as we are using degenerate type-II SPDC throughout) and the two terms inside the bracket of the Eq.\ref{tamp_qf_ana_eq} are nearly identical.
The oscillations in ETPA as function of $T_{e}$ in Fig.\ref{t6_spec_wptpte_fig}(d) are crucial to uncovering the transition paths as the transition amplitude in Eq.\ref{tamp_qf_ana_eq} includes the phase factor $\eta\=1-\e^{-\gamma_{e}T_{e}}\e^{-\i\Delta_{e}T_{e}}$. 
And therefore from the oscillations, one can identify all the responsible intermediate states and their energy $\varepsilon_{e}$ through $\Delta_{e}$. 
Fourier transforms of the signals $S_{\rm ETPA}(T_{e})$ are shown in Fig.\ref{t6_spec_wptpte_fig}(f) where we see that the peak position shifts toward the lower energy for $|f\rangle$ states from $|2\rangle$ to $|8\rangle$. 
These positions match well with the detunings for the intermediate state $|1\rangle$ and transitions follow the path $|0\rangle\rightarrow|1\rangle\rightarrow|f\rangle$.  
But the channel $|9\rangle$ does not follow this trend showing a peak at high energy (at 1.94 eV and higher than the channel $|2\rangle$) and therefore must be using other intermediate states.
To confirm this in Fig.\ref{t6_spec_wptpte_fig}(e) we have computed the same spectra as in Fig.\ref{t6_spec_wptpte_fig}(d) but dropped the intermediate states above $3.8$ eV (i.e., excluding the $|6\rangle$ and $|10\rangle$ intermediate states). 
Interestingly we notice that now the channel $|9\rangle$ disappears completely from the spectra even at $T_{e}\rightarrow 0$.
This is the direct evidence that channel $|9\rangle$ is not  populated through the intermediate state $|1\rangle$.
The peak position for the channel $|9\rangle$, Fig.\ref{t6_spec_wptpte_fig}(f) orange line, matches with the detunings for the intermediate state $|6\rangle$, and hence we conclude that it takes the path $|0\rangle\rightarrow|6\rangle\rightarrow|9\rangle$. 
Now we refer to Eq.\ref{tamp_qf_ana_te0_eq} where we have used the limit $T_{e}\rightarrow 0$.
The transition amplitude does not contain $\Delta_{e}$ terms in the denominator, resulting in a large amplitude. 
In other words, the broad signal and idler beams are generated in the limit $T_{e}\rightarrow 0$ and as a consequence highly detuned intermediate states (here $\Delta_{6}$ for the target state $|9\rangle$) become accessible.
At the same time, spectral resolution over the final excited states is not lost, as always the $\omega_{1}+\omega_{2}\=\omega_{p}$ condition is maintained in the SPDC process.
See the SI for the dependence of signal and idler spectral widths on $T_{e}$. 
Attaining this condition is impossible with classical light where a broad beam (generated through small $T_{p}$) cannot resolve different $|f\rangle$ states. \\ 

So far we have talked about the importance of entangled light at $T_{e}\rightarrow 0$ limit in exciting electronic states of molecules that are not accessible by classical light. 
It is of particular interest to harness the utility of entangled light in this limit and also for large $T_{e}$ values to excite each (bright) channel when excitation to the remaining states is suppressed, especially when the channels are very close in energy.
From now on we refer to this process with the term \textit{selectivity}.
In application to the T6 molecule, we select two excited states $|4\rangle$ and $|5\rangle$ to address selectivity as only these two excited states cannot be populated selectively with the one- and two-photon processes discussed above including ETPA even at $T_{e}\rightarrow 0$ limit.   
To proceed further beyond this limitation we optimize the two-photon state through the spectral phase and anticipate exciting each of the states $|4\rangle$ and $|5\rangle$, suppressing the population of close-lying states. 
Here we optimize the two-photon state of light by introducing a spectral phase $\phi(\omega)$ into the signal beam with two goals (a) enhancement of TPA and ETPA intensities, and (b) selective excitation of each channel (here states $|4\rangle$ and $|5\rangle$), assuming that the optimized phase can be generated through the use of SLM. \\

To generate flexible shapes for $\phi(\omega)$ used in Eq.\ref{field_eq} we model it as 
\begin{align}
    \label{phase_eq}
    \phi(\omega)\=\sum_{n=0}^{N_{p}}\alpha_{n}\mathcal{H}_{n}(\omega)\e^{-\frac{T_{p}^{2}}{8\ln 2}(\omega-\bar{\omega})^2},
\end{align}
where $\mathcal{H}_{n}$ are normalized $n$th order Hermite polynomials and $\alpha_{n}$ are expansion coefficients that we choose to vary in the range [-1, 1] with $N_{p}\=7$. 
In the experiment to realize flexible forms of $\phi(\omega)$ one can introduce computer-controlled discrete phase masks using SLM at different frequency components of entire spectral band\cite{Meshulach1998Coherent}. 
Clearly, the coefficients are free parameters and must be optimized within the specified range to obtain a suitable shape of $\phi(\omega)$ so that it can excite each channel selectively and enhance the absorption intensities.   
To quantify the enhancement and selectivity (for the target state $|f\rangle$) we define the following reward functions ($J_{t(f)}\equiv J_{t(f)}[\{\alpha_{t(f)}\}]$)
\begin{align}
    \label{reward_js_eq}
    J_{t}&\=\sum_{n\=1}^{N_{s}}P_{n}, \\
    \label{reward_jf_eq}
    J_{f}&\=\bar{P}_{f}-\sum_{n\neq f}^{N_{s}}\bar{P}_{n},
\end{align}
where $\bar{P}_{f}$ are normalized populations, $\sum_{n=1}^{N_{s}}\bar{P}_{n}=1$, with the number of excited states $N_{s}$.
Eq.\ref{reward_jf_eq} contains two terms, the target state population $\bar{P}_{f}$ and a penalty term $\widetilde{P}_{f}\=\sum_{n\neq f}\bar{P}_{n}$ for the rest of the populations.
We normalize the populations of possible excited states ($\bar{P}_{f}+\widetilde{P}_{f}\=1$), in order to compare the relative populations and take selectivity into account. 
The goal is to maximize the above reward functions by optimizing the coefficients in Eq.\ref{phase_eq}.  
For $N_{p}=7$ in Eq.\ref{phase_eq}, each of these optimization problems lives in an 8-dimensional hyperparameter space.
And the task is to locate the global maximum using an efficient optimization method as the evaluation of the reward function at each point is moderately expensive with a sufficiently large parameter space.  
A Gaussian process-based Bayesian optimization technique is used to maximize the above reward functions $J_{t}$ and $J_{f}$ exhausting the parameter space $\{\alpha\}$.  
We have implemented this in our problem using a Bayesian optimization solver\cite{Nogueira2014Bayesian}; see the SI for the details of the optimization.
We note that the mean photon numbers (with number operator $\hat{n}\=\int\d\omega \hat{E}^{\dagger}(\omega)\hat{E}(\omega)$) are constrained to a fixed value during the optimization. \\

\begin{figure}[h]
    \centering
    \includegraphics[width = 0.4 \textwidth]{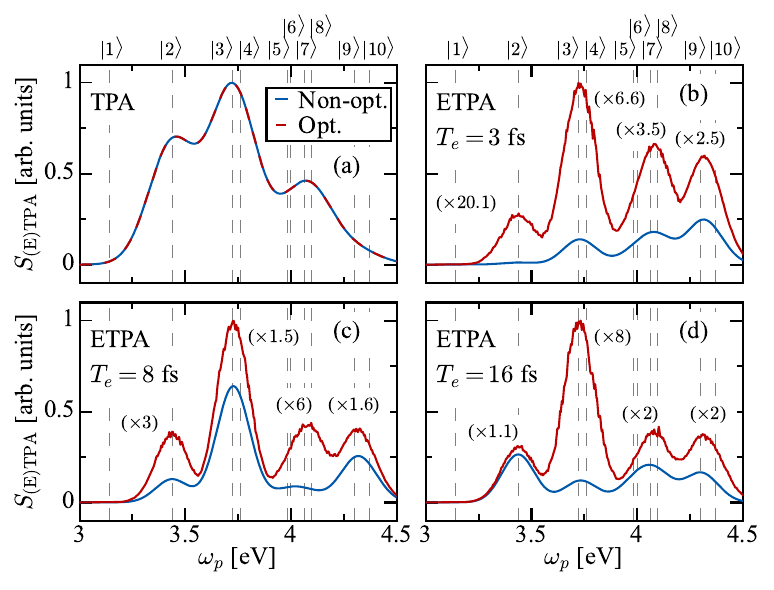}
    \caption{Optimized (a) TPA and (b-d) ETPA for three $T_{e}$ values: $3$, $8$, and $16$ fs. 
    Enhancement factors for bright channels are shown inside the brackets next to each peak.  
    }
   \label{t6_tot_wp_fig}
\end{figure}

TPA and ETPA intensities are maximized by optimizing the reward function $J_{t}$, Eq.\ref{reward_js_eq}, which is bounded between [0, 1]. 
We begin by optimizing TPA, where the optimized result is shown in Fig.\ref{t6_tot_wp_fig}(a). 
It turns out that the Fourier limited pulse always provides the maximum absorption strength i.e. when the spectral phase is independent of frequency.
As a result, spectral phase optimization does not improve the TPA strength. 
We note here that it is the highly detuned intermediate states (off-resonant) that are responsible for this result; resonant TPA is optimizable\cite{Dudovich2001Transform,Taro2002Optimization}.  
On the other hand, ETPA shows significant enhancements through the spectral phase optimization, Figs.\ref{t6_tot_wp_fig}(b-d).  
We considered three $T_{e}$ values for ETPA: 3 fs, 8 fs, and 16 fs.
All peaks (here $|2\rangle$, $|3\rangle$, $|8\rangle$, and $|9\rangle$) are amplified by a factor up to 20.
It is interesting to notice that for $T_{e}\=3$ fs, the lowest energy peak (channel $|2\rangle$) is not visible in the non-optimized ETPA spectrum but the optimization makes it prominent with a large amplitude.  
Similarly when $T_{e}=16$ fs (8 fs) is used, the optimized ETPA spectrum develops a distinct peak for channel $|3\rangle$ ($|8\rangle$) whereas the non-optimized ETPA appears with a little amplitude. 
We note that ETPA is also optimizable for a large $T_{e}$ value.
This optimization is robust and works for other molecules as well. 
We refer to the SI for ETPA optimization of a fused polythiophene 11 molecule and large $T_{e}$ values.  \\

To address the selective excitation of each channel we optimize $J_{f}$, Eq.\ref{reward_jf_eq}. 
The first term in this equation maximizes the populations of the $|f\rangle$ state whereas the second term is responsible for suppression of populations of the remaining states.  
A positive $J_{f}$ signifies that the $|f\rangle$ state is excited suppressing the populations of all other possible states and vice versa when it is negative. 
We note that it is bounded between [-1, 1].
$J_{f}\=+1$ ($J_{f}\=-1$) indicates the best (worst) selectivity and $J_{f}\=0$ when $\bar{P}_{f}$ equals to the penalty term $\widetilde{P}_{f}$. \\

\begin{figure}[h]
    \centering
    \includegraphics[width = 0.4 \textwidth]{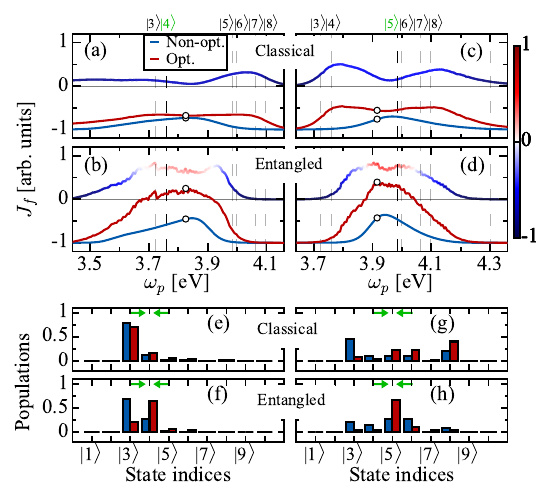}
    \caption{Reward functions for selective excitation of channels $|4\rangle$, and $|5\rangle$ with (a, c) classical and (b, d) entangled lights.
    The difference between the optimized and non-optimized reward functions is shown by the color density line where the color gradient dictates the selectivity -- blue for negative and red for positive selectivities. 
    The distribution of populations among 10 excited states for (e, g) classical and (f, h) entangled lights for selected points (shown by circles) in the corresponding reward function plots. 
    Target states are shown by green labels (for reward functions) and arrows (for distributions). 
    Here $T_{p}=10$ fs ($\sigma_{p}=0.18$ eV) and $T_{e}=1$ fs.
    }
   \label{t6_selec_wp_fig}
\end{figure}

\begin{figure}[h]
    \centering
    \includegraphics[width = 0.4 \textwidth]{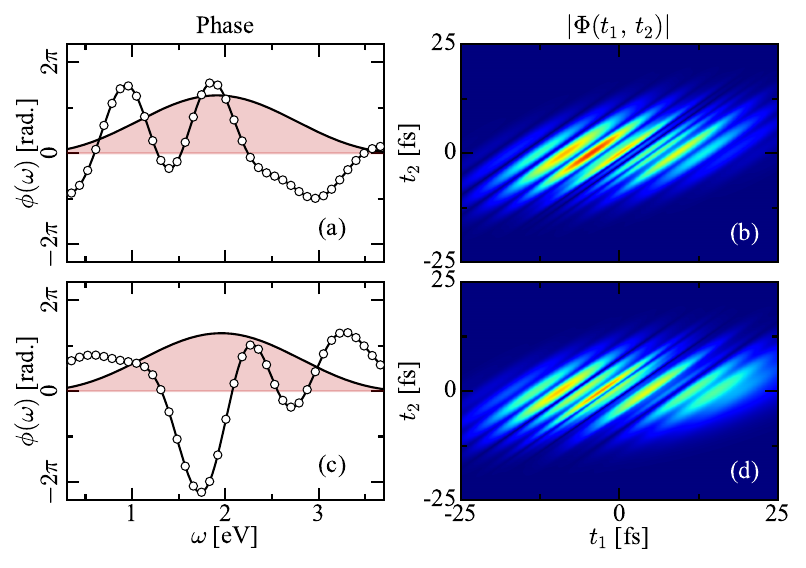}
    \caption{Spectral phase $\phi(\omega)$ (line with circles) and two-photon detection amplitude $|\Phi(t_{1},\,t_{2})|$ (color density plots) of optimized entangled light used to generate distributions (red) in Figs.\ref{t6_selec_wp_fig}(f, h).
    Here target states are (a, b) $|4\rangle$, and (c, d) $|5\rangle$.
    In the background of the spectral phase, the frequency content of the signal beam is shown by the red shaded area.  
    }
   \label{t6_selec_phase_fig}
\end{figure}

In Fig.\ref{t6_selec_wp_fig}, we compare the optimized $J_{f}$ with the non-optimized $J_{f}$ for both classical and entangled lights.  
We perform optimization for two target states $|f\rangle$: $|4\rangle$ and $|5\rangle$. 
As mentioned earlier, except for these two states the remaining states can be populated selectively using OPA, TPA, and ETPA.
Now we investigate whether the optimization method enables us to go beyond this limit, populating each of these two states selectively.  
We observe that the non-optimized $J_{f}$ for both TPA and ETPA is negative even around the resonance frequency and selectivity cannot be attained. 
In all the examples, the optimized two-photon states provide better selectivity (larger $J_{f}$) over a larger frequency range than the non-optimized states. 
Most notably a positive $J_{f}$ is obtained for the target states $|4\rangle$ and $|5\rangle$ only with optimized entangled light but always remains negative for all other situations including ETPA and optimized TPA.
The frequency range over which the optimized ETPA returns positive $J_{f}$ values is shown by red color gradients of the color density plots in Figs.\ref{t6_selec_wp_fig}(a-d).
Where we observe that the extension of the red region follows the skewness of the signal.  
For the target state $|4\rangle$, it is extended toward the high energy of the resonance line, whereas it is completely opposite for the target state $|5\rangle$. \\

The population distributions are shown in Figs.\ref{t6_selec_wp_fig}(e-h) for selected points in the corresponding reward function plots in Figs.\ref{t6_selec_wp_fig}(a-d) where the optimized ETPA returns positive $J_{f}$ values.
In the distributions, we see that for optimized light, either target state populations increase or penalty contributions decrease or both take place simultaneously compared to the non-optimized light.
Also, we notice that for the target state $|4\rangle$, non-optimized entangled light mostly excites the channel $|3\rangle$ with a probability 0.7.
But after the optimization, the dominating channel $|3\rangle$ is blocked and most of the population goes to the state $|4\rangle$ with a probability 0.65. 
Similarly, for the target channel $|5\rangle$, the non-optimized entangled light populates many excited states, mostly channels $|3\rangle$, $|4\rangle$, $|5\rangle$, and $|6\rangle$ (each channel with a probability of $>0.2$), and again optimized light successfully suppresses the population of all unwanted dominating channels (channels $|3\rangle$, $|4\rangle$, and $|6\rangle$) maximizing the transitions to channel $|5\rangle$ (with the probability 0.7). 
Further, we notice that optimization of classical light fails to return a positive $J_{f}$ for these two target states ($|4\rangle$ and $|5\rangle$) though it increases the selectivity after the optimization. 
For target state $|4\rangle$, both optimized and non-optimized light mostly populate $|3\rangle$ whereas for the target state $|5\rangle$ populations are distributed among several states, see Figs.\ref{t6_selec_wp_fig} (e and g). \\

Finally, in Table 1 we have summarized optical processes which selectively excite different excited states of the T6 molecule. 
We do not consider state $|10\rangle$ here since it is the highest energy level considered in this study and lacks neighboring states with even higher energies that are needed to direct selectivity for that channel.
Additionally, state $|7\rangle$, which turns out to be dark (in OPA, TPA, and ETPA) with zero transition dipoles, was not taken into account.
Optimized phase and two-photon detection amplitude $\Phi(t_{1},\,t_{2})$ are shown in Fig.\ref{t6_selec_phase_fig}.
The phases get structured around the spectral region of the signal beam and the detection amplitude stretches showing multiple lobes in time. 
There is a phase shift between the processes when the signal photon arrives earlier than the idler and the idler photon arrives earlier than the signal photon. 
Interference between transitions from different lobes of entangled light accessing detuned intermediate states acts constructively or destructively, providing selectivity over molecular electronic states. \\

\begin{table}
    \begin{tabular}{| c | c |} 
         \hline
         Target states & Optical processes  \\
         \hline\hline
         $|1\rangle$, $|6\rangle$ & OPA \\
         $|2\rangle$, $|3\rangle$, $|8\rangle$ & TPA, ETPA \\
         $|9\rangle$ & ETPA \\
         $|4\rangle$, $|5\rangle$ & O-ETPA \\
         \hline
    \end{tabular}
    \caption{Optical Processes Enabling Selective Excitation of Different Channels. 
    Here O-ETPA Stands for Optimized Entangled Two-Photon Absorption.  
    }
\end{table}

In conclusion, we have introduced an efficient two-photon optimization method using a Gaussian process based Bayesian optimization that outperforms standard TPA and ETPA in terms of enhanced absorption, spectral resolution, and selective excitation over molecular electronic states.
It has been found that our optimization method is sensitive to the unique temporal and spectral correlations of photon pairs.  
Therefore the usefulness of the optimization only arises with entangled light.  
We have reported that the optimized ETPA signal can be enhanced by several orders of magnitude with this approach and hence a lower photon flux can be used in imaging molecules that is ideal for preventing photodamage.
Also, this approach provides an important prediction for improving absorption intensities for measurements that are often quite challenging due to the low photon fluxes generated by SPDC sources. 
The optimization enables the selective excitation of electronic states that are very close in energy, which would otherwise be impossible with standard TPA and ETPA.
Along with the optimization protocol, we present mechanistic insights into comparing transition paths induced by entangled and classical light in the limit $T_{e}\rightarrow 0$, which have a direct impact on the resulting two opposite selectivities.
The spectral correlation in entangled light makes the highly detuned intermediate states accessible while also providing high resolution over two-photon states which is not possible with classical light.
The methods and principles addressed here are quite general and can also be applied to other organic chromophores with more complex electronic states where entanglement in the two-photon state and its optimization play a crucial role. 

\section{ASSOCIATED CONTENT}
The Supporting Information is available free of charge at \\

Technical details for calculations of two-photon transition amplitudes, T6 molecule, frequency content of SPDC photons, calculations of spectra in time and frequency domains, and Bayesian optimization are provided.

\section{AUTHOR INFORMATION}
\subsection{Corresponding Author}
\tb{George C. Schatz}\,--\,Department of Chemistry, Northwestern University, 2145 Sheridan Road, Evanston, Illinois 60208, United States; \\
Email: \tb{g-schatz@northwestern.edu}
\subsection{Author}
\tb{Sajal Kumar Giri}\,--\,Department of Chemistry, Northwestern University, 2145 Sheridan Road, Evanston, Illinois 60208, United States

\section{ACKNOWLEDGMENTS}
This research was supported by the Department of Energy, Office of Science, Biological and Environmental Research Program, under grant DE-SC0022118.  We thank Ted Goodson for helpful discussion.

\bibliography{draft.bib}

\end{document}

\begin{center}\bfseries\Large
Supporting Information: \\
Manipulating Two-Photon Absorption of Molecules through Efficient Optimization of Entangled Light
\\ [7mm]
\rm\large
Sajal Kumar Giri and 
George C. Schatz \\ [2mm]
\small{Department of Chemistry, Northwestern University, 2145 Sheridan Rd., Evanston, Illinois 60208, United States}
\end{center}

\vspace{0.2cm}

%\begin{quote}
%Technical details for calculations of two-photon transition amplitudes, T6 molecule, frequency content of SPDC photons, calculations of spectra in time and frequency domains, and Bayesian optimization are provided.
%Parameters for the numerical calculations are specified.
%\end{quote}

\section*{Contents}
\begin{itemize}
    \item[1.] Two-photon transition amplitudes 
    \item[2.] T6 molecule: Excited state energies and transition moments
    \item[3.] Frequency content of signal and idler beams 
    \item[4.] Time and frequency domain spectra
    \item[5.] Bayesian optimization
    \item[6.] ETPA Enhancement
\end{itemize}
\clearpage

\section{Two-photon transition amplitudes}
During the spontaneous parametric down-conversion (SPDC) process a pump pulse with frequency $\omega_{p}$ splits into two lower frequency photons historically known as signal and idler with frequencies $\omega_{1}$ and $\omega_{2}$ respectively.
The generated photons are divided into two categories based on their mutual polarization: type-I for parallel polarization and type-II for perpendicular polarization.
We have used degenerate type-II SPDC process in our work. 
The entangled two-photon state $|\Psi\rangle$ generated through the SPDC process is given by\cite{Fei1992Entanglement,Rubin1994Theory}
\begin{align}
  \label{tps_eq}
  |\Psi\rangle &\= \int_{-\infty}^{\infty}\d\omega_{2}\int_{-\infty}^{\infty}\d\omega_{1}\mathcal{J}(\omega_{1},\,\omega_{2})\e^{\i\phi(\omega_{1})}\hat{a}_{1}^{\dagger}(\omega_{1})\hat{a}_{2}^{\dagger}(\omega_{2})|0\rangle,
\end{align}
with the joint spectral amplitude (JSA) of detecting the signal photon with frequency $\omega_{1}$ and the idler photon with frequency $\omega_{2}$
\begin{align}
    \label{jsa_eq}
    \mathcal{J}(\omega_{1},\,\omega_{2})&\=\mathcal{N}\e^{-\frac{T_{p}^{2}}{8\ln 2}(\omega_{1}+\omega_{2}-\omega_{p})^{2}}\sinc(T_{e}[\omega_{1}-\omega_{2}]/2),
\end{align}
where $|0\rangle$ is the vacuum state of light, and $\sinc(x)\=\sin(x)/x$. 
$\mathcal{J}$ includes the pump pulse amplitude and phase matching function. 
The pump pulse duration is controlled with $T_{p}$ which is inversely proportional to its spectral width $\sigma_{p}$, where $\sigma_{p}\=4\ln 2/T_{p}$. 
The phase matching function comes with the entanglement time $T_{e}$ that can be controlled with the crystal length in SPDC (modifying the travel time difference between the photons inside the nonlinear crystal). 
In Eq.\ref{tps_eq} $\phi(\omega_{1})$ is the spectral phase for the signal beam. 
Clearly one can see that $|\Psi\rangle$ in Eq.\ref{tps_eq} can not be expressed as a direct product of two frequency components and hence they are entangled, $|\Psi\rangle\,\neq\,|f(\omega_{1})\rangle\otimes |g(\omega_{2})\rangle$. 
Characteristics of the entangled two-photon state mostly rely on energy conservation ($\omega_{p}\=\omega_{1}+\omega_{2}$) and momentum conservation ($\mb{k}_{p}(\omega_{p})\=\mb{k}_{1}(\omega_{1})+\mb{k}_{2}(\omega_{2})$).
Unlike classical light, for a short $T_{e}$ the bandwidth of the signal and idler beams can be much wider than the pump pulse, but together they deliver a limited spectral width (independent of the bandwidth
of individual photons) because $\omega_{p}\=\omega_{1}+\omega_{2}$ is always maintained.
This feature of entangled light has a dramatic significance in exciting molecular electronic states following some specific paths not accessible by classical light.
The factor $\mathcal{N}$ in Eq.\ref{jsa_eq} normalizes $\mathcal{J}$ so that $\int\int\d\omega_{1}\d\omega_{2}\mathcal{J}(\omega_{1},\,\omega_{2})=1$ is always maintained. \\

\begin{figure}[h]
    \centering
    \includegraphics[width = 0.4 \textwidth]{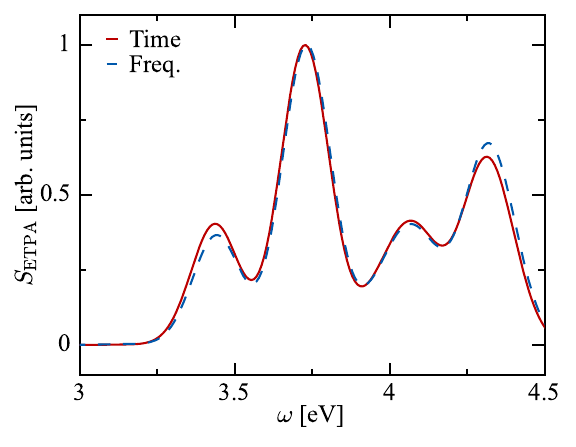}
    \caption{ETPA spectra calculated in time (red) and frequency (blue) domains. 
    For frequency domain spectra we have used the narrow band approximation. 
    Here, $T_{p}=10$ fs, and $T_{e}=1$ fs.
    }
   \label{pop_ana_wp_fig}
\end{figure}

In order to obtain two-photon absorption spectra, we compute dipole and field correlation functions numerically by solving Eqs.(1 and 2) respectively in the main text without invoking any additional approximations.
Note that we derived the transition amplitudes in the frequency domain using the narrow band approximation of the signal and idler beams (i.e., $\sigma_{p} << \bar{\omega}_{1,\,2}$).  
Also, a closed analytical expression for the transition amplitude cannot be obtained for an arbitrary form of the spectral phase $\phi(\omega)$ (which we have considered in optimizing the two-photon state).
For a sufficiently small $\sigma_{p}$, we find an excellent agreement between the time and frequency domain spectra, see Fig.\ref{pop_ana_wp_fig}. \\

Using a narrow band approximation of the generated photons ($\sigma_{p}<<\bar{\omega}_{1,\,2}$) we can safely approximate the electric field operator as 
\begin{align}
    \hat{E}_{n}^{\dagger}(t)&\=\sqrt{\frac{\bar{\omega}_{n}}{cA}}\e^{-\i\bar{\omega}_{n}t}\int_{-\infty}^{\infty}\d\omega_n\hat{a}_{n}(\omega_{n}),
\end{align}
and with this the field correlation function $\Phi(t_{1},\,t_{2})\=$ $\langle 0|\hat{E}_{1}^{\dagger}(t_{1})\hat{E}_{2}^{\dagger}(t_{2})|\Psi\rangle$ is obtained as 
\begin{align}
    \Phi(t_{1},\,t_{2})&\,\propto\,\sqrt{\bar{\omega}_{1}\bar{\omega}_{2}}\e^{-\i(\bar{\omega}_{1}t_{1}+\bar{\omega}_{2}t_{2})}\int_{-\infty}^{\infty}\d\omega_{1}\int_{-\infty}^{\infty}\d\omega_{2}\e^{\i\phi(\omega_{1})}\e^{-\frac{T_{p}^{2}}{8\ln 2}(\omega_{1}+\omega_{2}-\omega_{p})^{2}}\sinc\Bigg{(}\frac{T_{e}(\omega_{1}-\omega_{2})}{2}\Bigg{)} \\
    &\overset{\phi\rightarrow 0}{\=}\sqrt{\frac{T_{p}\bar{\omega}_{1}\bar{\omega}_{2}}{T_{e}}}\e^{-\i\bar{\omega}_{1}t_{1}}\e^{-\i\bar{\omega}_{2}t_{2}}\e^{-\frac{2\ln2}{T_{p}^{2}}(t_{1}+t_{2})^{2}/4}\Pi\Bigg{(}\frac{t_{1}-t_{2}}{2T_{e}}\Bigg{)},
\end{align}
where $\Pi(x)$ is a trapezoidal function, $\Pi(x)\=1/2$ for $|x|<1/2$ and otherwise $\Pi(x)=0$.
Similarly we calculate $\langle 0|\hat{E}_{2}^{\dagger}(t_{1})\hat{E}_{1}^{\dagger}(t_{2})|\Psi\rangle$ and using them we obtain the transition amplitude as 
\begin{align}
    \label{tamp_qf_ana_eq}
    T_{f}^{ab}&\,\propto\,\sqrt{\frac{T_{p}\bar{\omega}_{1}\bar{\omega}_{2}}{T_{e}}}\e^{-\frac{T_{p}^{2}}{8\ln 2}(\varepsilon_{f}-\varepsilon_{g}-\omega_{p})^{2}}\sum_{e}D_{e}^{ab}\Bigg{[}\frac{1-\e^{-\gamma_{e}T_{e}/2}\e^{-\i\Delta_{e}^{(1)}T_{e}}}{\Delta_{e}^{(1)}-\i\gamma_{e}/2} +\frac{1-\e^{-\gamma_{e}T_{e}/2}\e^{-\i\Delta_{e}^{(2)}T_{e}}}{\Delta_{e}^{(2)}-\i\gamma_{e}/2}\Bigg{]},
\end{align}
with detuning of the intermediate states $\Delta_{e}^{(1,\,2)}=\varepsilon_{e}-\varepsilon_{g}-\bar{\omega}_{1,\,2}$.

\section{T6 molecule: Excited state energies and transition moments}
The optimized geometry shown in the main text has a non-planar geometry with twisted thiophene monomer units.
At room temperature, this molecule can undergo a dynamic evolution over different conformers including the planar shape.
It has been observed that the planar and twisted geometries have similar optical properties\cite{Kang2020Efficient}.  
In order to keep things simple and preserve the molecule's fundamental optical properties (one- and two-photon), we only take into account the optimized equilibrium geometry.
To check the Hermiticity of the dipole matrix we compare $\langle m|\hat{\mu}|n\rangle^{*}$ with $\langle n|\hat{\mu}|m\rangle$, where $m$ and $n$ are integers between 0 and 10. 
We find that $\langle m|\hat{\mu}|n\rangle^{*}$ and $\langle n|\hat{\mu}|m\rangle$ are very close but not exactly the same.
This is because the reference states in SLR-TDDFT are different for these two matrix elements.
$|m\rangle$ is the reference state in computing $\langle m|\hat{\mu}|n\rangle$ whereas $|n\rangle$ is the reference state when it computes $\langle n|\hat{\mu}|m\rangle$.   
To circumvent this small error we symmetrize it using only the lower diagonal matrix elements.   
In calculating the one-photon absorption (OPA) spectra, peaks are broadened using a Gaussian function with a full-width at half-maximum (FWHM) of 0.1 eV. \\

\section{Frequency content of signal and idler beams}

The spectral content of the signal and idler beams is calculated as\cite{Schlawin2012Manipulation} 
\begin{align}
    \label{spec_sig_eq}
    n(\omega)\,\propto\,\int\d\omega'|A_{p}(\omega+\omega')|^2\sinc^2(T_{e}(\omega-\omega')/2),
\end{align}
where $A_{p}$ is the spectral envelope of the pump pulse
\begin{align}
    A_{p}(\omega+\omega')=\e^{-\frac{T_{p}^{2}}{8\ln 2}(\omega+\omega'-\omega_{p})^{2}}.
\end{align}

\begin{figure}[h]
    \centering
    \includegraphics[width = 0.4 \textwidth]{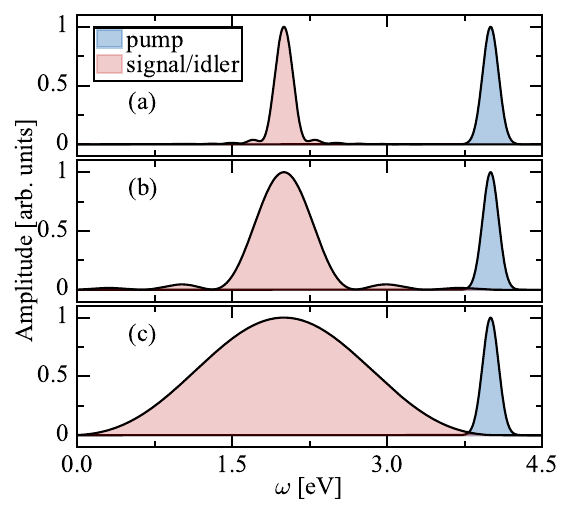}
    \caption{Spectral amplitudes (calculated through Eq.\ref{spec_sig_eq}) of signal beam (red) and pump pulse (blue) for (a) $T_{e}=10$ fs, (b) $T_{e}=3$ fs, and (c) $T_{e}=1$ fs. 
    We have considered identical signal and idler beams ($\bar{\omega}_{1}=\bar{\omega}_{2}$).
    Here $T_{p}=10$ fs ($\sigma_{p}=0.18$ eV).
    }
   \label{sig_idler_width_fig}
\end{figure}

\section{Time and frequency domain spectra}

\section{Bayesian optimization}
The Bayesian optimization (BO) method\cite{Garnett2022Bayesian} is focused on solving the following problem
\begin{align}
    {\rm max}_{\alpha}f(\alpha),
\end{align}
where $\alpha\,\in\,\mathbb{R}^{d}$ with $d\,\leq\,20$ and $f$ is a continuous function, $f:\mathbb{R}^{d}\rightarrow\mathbb{R}$.
Typically $f$ is modeled with Gaussian process regression. 
It is treated as a black box function expensive to evaluate where we do not have its analytical expression and also its derivatives are unknown. 
The goal is to locate $\alpha_{*}$ in a minimum number of steps that corresponds to a global rather than the local optimum. 
Let us define the observation data set up to $t$ steps with inputs $\boldsymbol{\alpha}\=\{\alpha^{(n)}\}_{1}^{t}$ and the corresponding target values $\mb{S}\=\{S^{(n)}\}_{1}^{t}$ as $\mathcal{O}_{t}\=\{(\alpha^{(1)},S^{(1)}),\ldots,(\alpha^{(t)},S^{(t)})\}$. 
The BO procedure can be summarized in the following steps: 
\begin{itemize}
    \item[1.] It begins with the use of a surrogate model approximating the true function $f$ called prior and an acquisition function. 
    \item[2.] Then find the hyperparameter $\alpha_{*}$ where the acquisition function is maximized. 
    The acquisition function depends on the prior function which decides where the next sample should be.   
    \item[3.] Evaluate the reward function at the new point $\alpha_{*}$ and generate the posterior using the Bayes rule
    \begin{align}
        p(w|\mathcal{O})\=\frac{p(\mathcal{O}|w)p(w)}{p(\mathcal{O})},
    \end{align}
    where $p(w|\mathcal{O})$ is the posterior probability, $p(w)$ is the prior distribution of parameters $w$, and $p(\mathcal{O}|w)$ is the likelihood function.  
    It includes information of both prior and observations. 
    \item[4.] This new evaluation adds to the old observations and the process is repeated through step 2 until convergence is reached.  
\end{itemize}

\subsection{Gaussian processes regression}
We use Gaussian process regression (GPR) as a prior in the BO. 
With the mean $m(\alpha)$ and covariance functions $k(\alpha,\,\alpha')$
\begin{align}
    m(\alpha)&\=\mathbb{E}[f(\alpha)] \\
    k(\alpha,\,\alpha')&\=\mathbb{E}[(f(\alpha)-m(\alpha))(f(\alpha')-m(\alpha'))]  
\end{align}
it can be expressed as 
\begin{align}
    f(\alpha)\,\sim\,\mathcal{GP}(m(\alpha),\,k(\alpha,\,\alpha')),
\end{align}
where $\mathcal{GP}$ denotes the Gaussian process.
The covariance function determines how the function's value at one point ($\alpha'$) influences the distribution of the function's value at other places ($\alpha$).
The covariance kernel function has several alternatives; as long as it adheres to the characteristics of a kernel, it can take many different forms.
We have considered the squared exponential covariance function 
\begin{equation}
    k(\alpha,\,\alpha')\=\exp\Big(-\frac{1}{2}|\alpha-\alpha'|^{2}\Big).
\end{equation}
The observation data and new points are jointly distributed
\begin{align}
    \begin{bmatrix}
        f \\
        f_{*}
    \end{bmatrix}
    \,\sim\,\mathcal{N}\Bigg(
    \begin{bmatrix}
        m(\alpha) \\
        m{(\alpha_{*})}
    \end{bmatrix},
    \begin{pmatrix}
        k(\alpha,\,\alpha) & k(\alpha,\,\alpha_{*}) \\
        k(\alpha_{*},\,\alpha) & k(\alpha_{*},\,\alpha_{*}) 
    \end{pmatrix}\Bigg).
\end{align}
$f_{*}$ can be obtained as 
\begin{align}
    f_{*}\,\sim\,\mathcal{N}(m(f_{*}),\,{\rm cov}(f_{*})),
\end{align}
where the mean $m(f_{*})$ and covariance ${\rm cov}(f_{*})$ matrices are given by
\begin{align}
    m(f_{*})&\=m(\alpha_{*})+k(\alpha_{*},\,\alpha)k(\alpha,\,\alpha)^{-1}f, \\
    {\rm cov}(f_{*})&\=k(\alpha_{*},\,\alpha_{*})-k(\alpha_{*},\,\alpha)k(\alpha,\,\alpha)^{-1}k(\alpha,\,\alpha_{*}).
\end{align}

\subsection{Acquisition functions}
Acquisition functions in the BO play a significant role which decides the sampling in the hyperparameter space.
They keep the balance between exploration and exploitation.
Exploitation refers to sampling in areas with a high objective and high prediction uncertainty, whereas exploration refers to sampling in areas with a low objective and high prediction uncertainty.
There are several types of acquisition functions available and each of them has advantages and disadvantages. 
We decided to use expected improvement which is widely used in many applications.
In the expected improvement, the sampling point at $(t+1)$th step is given by
\begin{align}
    \alpha_{t+1}\={\rm argmax}_{\alpha}\mathbb{E}({\rm max}\{0, h_{t+1}(\alpha)-f(\alpha^{+})\},\,\mathcal{O}_{t}),
\end{align}
where $h_{t+1}$ is the posterior mean at $(t+1)$th step, $f(\alpha^{+})$ is the maximum value obtained until the step $t$. 

\begin{figure}[h]
    \centering
    \includegraphics[width = 0.85 \textwidth]{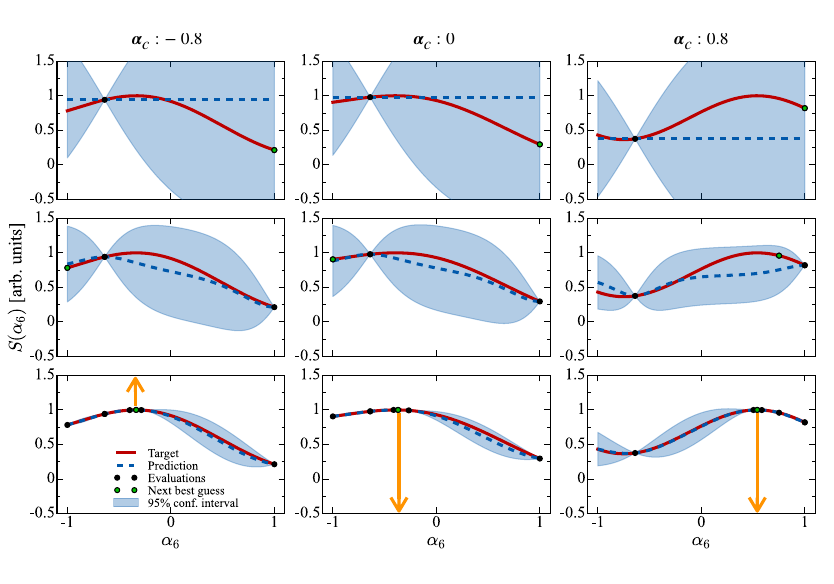}
    \caption{Examples of phase parameter optimization ($\alpha_{6}$) for ETPA spectra with three cuts in the hyperparameter space: All parameters except $\alpha_{6}$ are set to (Left) -0.8, (Middle) 0, and (Right) 0.8. 
    The number of iterations increases from top to bottom.  
    }
   \label{gp_fig}
\end{figure}

\section{ETPA enhancement}

\begin{figure}[h]
    \centering
    \includegraphics[width = 0.5 \textwidth]{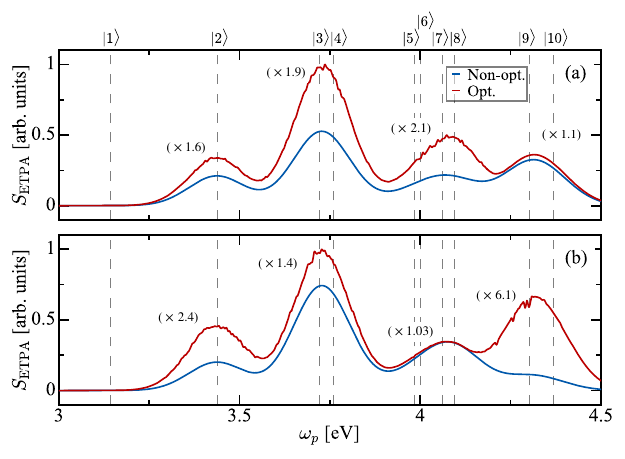}
    \caption{Optimized ETPA for T6 molecule for (a) $T_{e}=1$ fs and (b) $T_{e}=2$ fs.
    Enhancement factors for bright channels are shown inside the brackets next to each peak.
    }
   \label{t6_tot_wp_2_te_fig}
\end{figure}

\begin{figure}[h]
    \centering
    \includegraphics[width = 0.5 \textwidth]{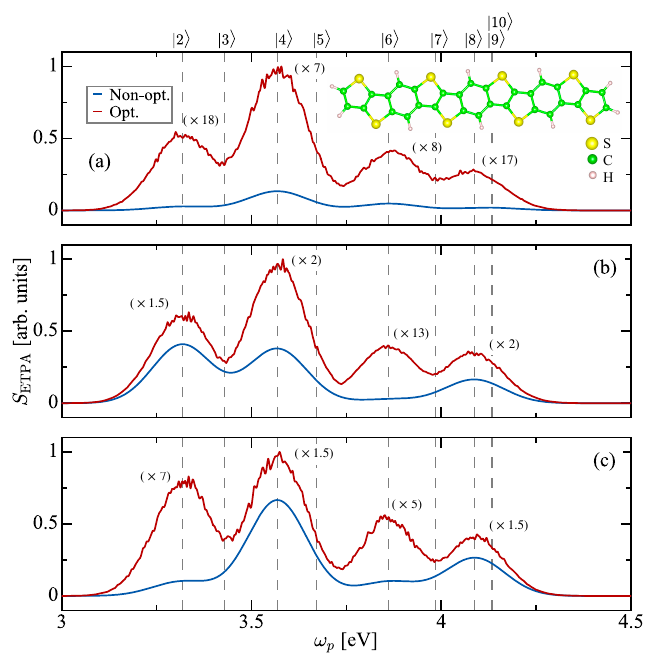}
    \caption{Optimized ETPA for fused polythiophene 11 molecule for (a) $T_{e}=3$ fs, (b) $T_{e}=8$ fs and (c) $T_{e}=16$ fs. 
    Enhancement factors for bright channels are shown inside the brackets next to each peak.
    The DFT optimized geometry of the molecule is shown in the inset of panel (a). 
    }
   \label{polythiophene11_tot_wp_2_te_fig}
\end{figure}

Like T6 molecule we optimize and calculate the first 10 excited states for a fused polythiophene 11 molecule, inset of the Fig.\ref{polythiophene11_tot_wp_2_te_fig}(a). 
Following the same procedure described in the main text we optimize the ETPA for this molecule. 
The optimized and non-optimized ETPA spectra are compared in Fig.\ref{polythiophene11_tot_wp_2_te_fig}. 

\clearpage

%\printbibliography

\bibliographystyle{unsrt}
\bibliography{supp}